\documentclass[twocolumn,showpacs,preprintnumbers,amsmath,amssymb]{revtex4}

\usepackage{graphicx}
\usepackage{dcolumn}
\usepackage{bm}

\begin{document}


\title{Electric-field control of tunneling magnetoresistance effect in a Ni/InAs/Ni quantum-dot spin valve}

\author{K. Hamaya,\footnote{hamaya@iis.u-tokyo.ac.jp} M. Kitabatake, K. Shibata, M. Jung, M. Kawamura, K. Hirakawa,\footnote{Also at: Collaborative Institute for Nano Quantum Information Electronics, University of Tokyo.} and T. Machida$^{\dag}$\footnote{tmachida@iis.u-tokyo.ac.jp}}
\affiliation{%
Institute of Industrial Science, University of Tokyo, 4-6-1 Komaba, Meguro-ku, Tokyo 153-8505, Japan
}%

\author{T. Taniyama}
\affiliation{%
Materials and Structures Laboratory, Tokyo Institute of Technology, 4259 Nagatsuta, Midori-ku, Yokohama 226-8503, Japan 
}%

\author{S. Ishida and Y. Arakawa$^{\dag}$}
\affiliation{%
Research Center for Advanced Science and Technology and Institute of Industrial Science, University of Tokyo, 4-6-1 Komaba, Meguro-ku, Tokyo 153-8505, Japan }

%

\date{\today}
\begin{abstract}
We demonstrate an electric-field control of tunneling magnetoresistance (TMR) effect in a semiconductor quantum-dot spin-valve device. By using ferromagnetic Ni nano-gap electrodes, we observe the Coulomb blockade oscillations at a small bias voltage. In the vicinity of the Coulomb blockade peak, the TMR effect is significantly modulated and even its sign is switched by changing the gate voltage, where the sign of the TMR value changes at the resonant condition.
\end{abstract}
\maketitle
For future spintronic applications such as a spin field-effect transistor (spin-FET),\cite{Datta,Schliemann} the injection of spin-polarized electrons, manipulation of the injected spins, and detection of the manipulated spins are technological requirements.\cite{Awschalom} Using carbon nanotubes (CNTs) as a one- or zero-dimensional conduction channel, the electric-field control of spin transport was recently achieved.\cite{Sahoo,Nagabhirava,Jensen,Man} However, there has been no clear experimental demonstration of the above three requirements in a single ferromagnet/semiconductor hybrid device. As compared to the CNTs, semiconductor systems are scalable and compatible with existing nano-fabrication technologies. Also, semiconductor-based quantum-dot (QD) systems have controllable characteristics for the number of electrons, orbital states, electron-spin and nuclear-spin states.\cite{Fujisawa,Ono1,Koppens}

Very recently, we observed spin transport through a single semiconductor QD using Co/InAs QD/Co lateral spin-valve devices.\cite{Hamaya} The spin transport is likely to be based on spin accumulation in the QD.\cite{Barnas,Imamura,Brataas,Jan,Weymann} However, the coupling of electrons between the InAs QD and the Co electrodes was very weak, so that evident Coulomb oscillation peaks could not be gained: the spin transport was observed only in a nonlinear transport regime under a large bias condition. As a consequence, we could not realize a systematic gate-control of spin transport, and did not understand correlation between the Coulomb blockade characteristics and the observed spin transport features.

In this letter, we demonstrate an electrical control of spin transport in a Ni/InAs/Ni QD spin-valve device. By using Ni electrodes, we improve the Coulomb blockade characteristics compared to our previous results for Co electrodes.\cite{Hamaya} Near the Coulomb blockade peak, the TMR effect is significantly varied and even its sign is switched by changing the gate voltage, where the sign of the TMR value changes at the resonant condition. This is the first demonstration of gate-tunable TMR through a semiconductor channel.
\begin{figure}[t]
\includegraphics[width=8.5cm]{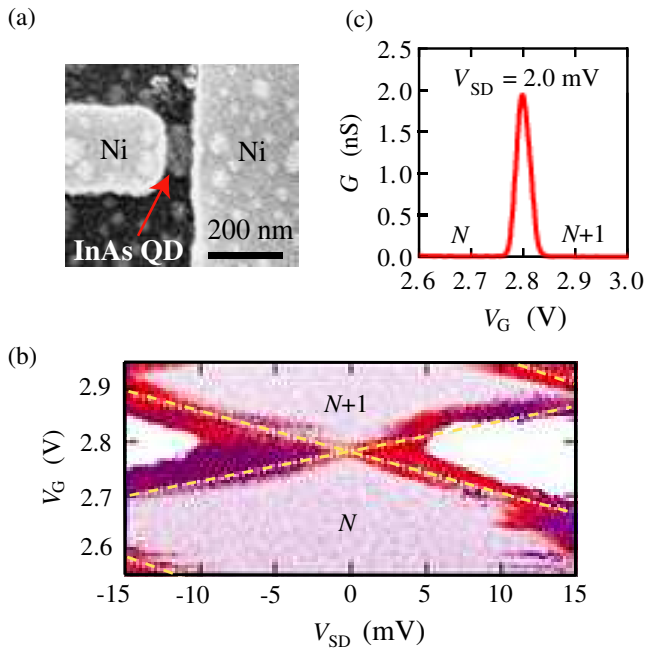}
\caption{(a) A scanning electron micrograph of our Ni/InAs/Ni quantum-dot spin-valve device. (b) Differential conductance, $dI$/$d$$V$$_\mathrm{SD}$, as a function of $V$$_\mathrm{SD}$ and $V$$_\mathrm{G}$ at 0.5 K. The regions enclosed by the dashed line show Coulomb blockade regime. The two white areas outside the enclosed regions are beyond the limits of measurement range. (c) The conductance as a function of $V$$_\mathrm{G}$ measured at $V$$_\mathrm{SD} =$ 2 mV.}
\end{figure}  
 
We grew self-assembled InAs QDs on a substrate consisting of 170-nm-thick GaAs buffer layer/90-nm-thick AlGaAs insulating layer/$n$$^{+}$-GaAs(001). The $n$$^{+}$-GaAs(001) was used as a backgate electrode. The wire-shape Ni electrodes with a $\sim$ 70-nm gap were fabricated by using an electron-beam lithography and a lift-off method. Before evaporation of Ni thin film, we etched InAs surface with buffered HF solution. To induce asymmetric shape anisotropy, the Ni wires were designed to be $\sim$ 0.25 $\mu$m wide and 20 $\mu$m long and $\sim$ 0.75 $\mu$m wide and 20 $\mu$m long. As shown in Fig. 1(a), a single InAs QD ($\sim$ 120 nm) is in contact with two Ni wires which have a thickness of 40 nm: this is a lateral QD spin-valve device. A schematic diagram of a QD spin-valve device was described in our previous report.\cite{Hamaya} Transport measurements were performed by the dc method at $\sim$ 0.5 K. External magnetic fields ($H$) were applied parallel to the long axis of the wire-shape ferromagnetic electrodes (FEs), and the sweep rate was less than 0.001 T/s. 

Figure 1(b) shows the differential conductance, $dI$/$d$$V$$_\mathrm{SD}$, as a function of $V$$_\mathrm{SD}$ and $V$$_\mathrm{G}$ under the parallel magnetic configuration of the FEs ($B \sim$0.1 T). Even for the FEs, we observe clear Coulomb diamonds. The charging energy is roughly estimated to be $\sim$ 15 meV, being consistent with our previous works using nonmagnetic Au electrodes.\cite{Jung} The conductance change versus $V$$_\mathrm{G}$ is also measured as shown in Fig. 1(c) at $V$$_\mathrm{SD} =$ 2 mV. A clear Coulomb oscillation peak is observed, where the number of electrons fluctuates between $N$ and $N +$ 1,\cite{ref} while the plateaus outside the peak represent the Coulomb blockade regime, where the number of electrons is fixed ($N$ or $N +$ 1). We note that such a Coulomb blockade oscillation peak is gained for the first time by using Ni electrodes instead of Co ones.\cite{Hamaya}

By applying external magnetic fields ($B$) parallel to the Ni wire axis, we measure field-dependent magnetoresistance (MR) for various $V$$_\mathrm{G}$ in a finite conductance regime near the Coulomb oscillation peak. At $V$$_\mathrm{G} =$ 2.78 and 2.79 V [Figs. 2(a) and 2(b)], evident hysteretic MR curves with a positive sign are gained. These data indicate spin transport via the QD separated by a double tunnel barrier.\cite{Hamaya} Due to the asymmetric wire shape, the difference in the magnetization switching fields between two FEs is induced. As a result, anti-parallel magnetic configuration of the FEs can be formed at around $B \sim$ 0.015 T [see - inset of Fig. 2(a)]. Surprisingly, the sign of the MR curves is switched to negative at $V$$_\mathrm{G} =$ 2.80 and 2.82 V  [Figs. 2(c) and 2(d)]. We note that the magnitude of the TMR is also varied markedly by changing $V$$_\mathrm{G}$.
\begin{figure}[t]
\includegraphics[width=8.5cm]{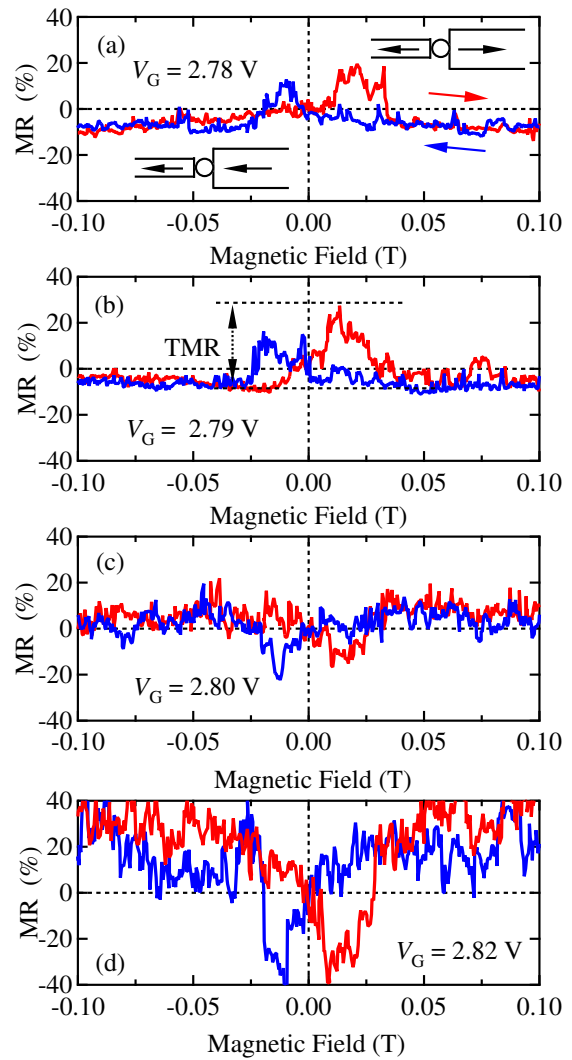}
\caption{TMR changes with $V$$_\mathrm{G}$. The MR ratio (\%) is denoted as $\{$($I_\mathrm{0} -$ $I_\mathrm{B}$)/$I_\mathrm{B}$$\}$ $\times$ 100, where $I_\mathrm{0}$ and $I_\mathrm{B}$ are the tunnel currents for zero-field and for a magnetic field of $B$, respectively. The TMR value is defined as the maximum change in the MR value showing hysteretic behavior  [see-(b)]. The gate voltages are $V$$_\mathrm{G} =$ (a) 2.78 V, (b) 2.79 V, (c) 2.80 V, and (d) 2.82 V. }
\end{figure}  

In Fig. 3 we summarize the TMR values as a function of $V$$_\mathrm{G}$ (red plots), together with the Coulomb blockade peak shown in Fig. 1(c). We can see the systematic variation in the TMR with applying $V$$_\mathrm{G}$, in which the TMR value is changed from positive ($\sim$ +30\%) to negative ($\sim -$60\%). This is the first demonstration of the sign controlled TMR by the gate voltage in ferromagnet/semiconductor hybrid devices. In addition, the TMR values are quite larger than that reported for CNT spin valves.\cite{Sahoo,Nagabhirava} Then, the magnitude of the TMR is large compared to the value of $\sim$ 21 \%, based on the simple Julliere model\cite{Julliere} for Ni electrodes (${\it P}_\mathrm{L} =$${\it P}_\mathrm{R} \sim$ 0.31, where ${\it P}_\mathrm{L}$ and ${\it P}_\mathrm{R}$ are the spin polarization of the density of states at the Fermi energy of the two FEs). It should be noted that the sign of the TMR changes at $V$$_\mathrm{G} \sim$ 2.80 V where the conductance peak is observed.
\begin{figure}[t]
\includegraphics[width=8.5cm]{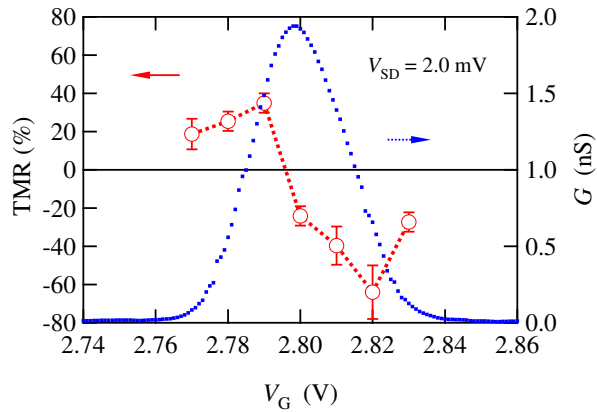}
\caption{The magnitude and sign changes of the TMR with $V$$_\mathrm{G}$ (left axis), together with the Coulomb oscillation data given in Fig 1(c) (right axis).}
\end{figure}   

A qualitative explanation of the sign changed TMR effect was described in previous report for CNT spin-valve devices.\cite{Sahoo,Cottet} For parallel (anti-parallel) magnetic configuration of the FEs, transmission probability $T$$^{\sigma}_\mathrm{P[AP]}$ of electrons through a QD near the resonance can be written by the spin-dependent Breit-Wigner-like formula, 
\begin{equation}
T^{\sigma}_\mathrm{P[AP]} = \frac{{\it \Gamma}^{\sigma}_\mathrm{L} {\it \Gamma}^{\sigma}_\mathrm{R}}{(E-E^{\sigma}_\mathrm{0})^{2} + ({\it \Gamma}^{\sigma}_\mathrm{L} + {\it \Gamma}^{\sigma}_\mathrm{R})^{2}/4},  
\end{equation}
where ${\it \Gamma}$$^{\sigma}_\mathrm{L[R]}$ is the spin-dependent coupling to the left (right) FEs, and $E$$^{\sigma}_\mathrm{0}$ is the spin-dependent energy level in the QD.\cite{Sahoo,Cottet} When $|$$E$$-$$E$$^{\sigma}_\mathrm{0}$$|$ $\gg$ (${\it \Gamma}$$^{\sigma}_\mathrm{L}$ $+$ ${\it \Gamma}$$^{\sigma}_\mathrm{R}$), i.e., off resonance, $T$$^{\sigma}_\mathrm{P[AP]}$ is proportional to ${\it \Gamma}$$^{\sigma}_\mathrm{L}$${\it \Gamma}$$^{\sigma}_\mathrm{R}$, resulting in the positive MR of  2${\it P}_\mathrm{L}$${\it P}_\mathrm{R}$/(1 $-$ ${\it P}_\mathrm{L}$${\it P}_\mathrm{R}$),  where the sign of ${\it P}_\mathrm{L}$ and ${\it P}_\mathrm{R}$ is the same. When $E$ $=$ $E$$^{\sigma}_\mathrm{0}$, i.e., at resonance, assuming a very asymmetric coupling of ${\it \Gamma}$$^{\sigma}_\mathrm{L}$ $\ll$ ${\it \Gamma}$$^{\sigma}_\mathrm{R}$, one can obtain $T$$^{\sigma}_\mathrm{P[AP]}$ $=$ 4${\it \Gamma}$$^{\sigma}_\mathrm{L}$/${\it \Gamma}$$^{\sigma}_\mathrm{R}$ from Eq. (1), leading to the negative MR of $-$2${\it P}_\mathrm{L}$${\it P}_\mathrm{R}$/(1 $+$ ${\it P}_\mathrm{L}$${\it P}_\mathrm{R}$). From these considerations, the gate control of the TMR presented here is basically interpreted in terms of the spin-dependent resonant tunneling model.\cite{Sahoo,Cottet} 

Using this QD spin-valve device, we can develop an all-electrical means for the injection of spin-polarized electrons, manipulation of the injected spins, and detection of the manipulated spins in a single device. The present work is the first demonstration of the gate-tunable non-volatile-memory effect for ferromagnet/semiconductor hybrid devices with a semiconductor channel. This becomes the first step toward the development of a semiconductor-based spin-FET where spin transport via a single-spin state can be controlled by the electric field and the magnetization direction of the FEs.



\begin{thebibliography}{11}
\bibitem{ref}
Although $N$ could not be assigned precisely, we speculate 10 $\lesssim$$N$$\lesssim$ 20 in the present work.
\end{thebibliography}
\end{document}